\documentclass{PoS}

\newcommand{\oG}{\overline{G}}
\newcommand{\p}{\partial}

\title{Infrared behavior of the gluon and ghost propagators in Yang-Mills
theories }

\ShortTitle{Infrared behavior of the gluon and ghost propagators in
Yang-Mills theories }

\author{\speaker{Silvio Paolo Sorella}\thanks{Work supported by FAPERJ, Funda{\c c}{\~a}o de Amparo
{\`a} Pesquisa do Estado do Rio de Janeiro, under the program {\it
Cientista do Nosso Estado}, E-26/151.947/2004. }, Marcio Capri,
Vitor Lemes, Rodrigo Sobreiro, Ronaldo Thibes\\UERJ - Universidade
do Estado do Rio de Janeiro, Rua S\~{a}o Francisco Xavier 524,
20550-013 Maracan\~{a}, Rio de Janeiro, Brasil\\        E-mail:
\email{sorella@uerj.br,marcio@dft.if.uerj.br,\\vitor@dft.if.uerj.br,sobreiro@dft.if.uerj.br,thibes@dft.if.uerj.br}}

\author{John Gracey\\Theoretical Physics Division,
Department of Mathematical Sciences, University of
Liverpool\footnote{Preprint LTH--728}, P.O. Box 147, Liverpool,
L69 3BX, United Kingdom\\E-mail: \email{jag@amtp.liv.ac.uk}}

\author{David Dudal\thanks{Postdoctoral fellow of the
\emph{Special Research Fund} of Ghent University.}, Henri
Verschelde\\Ghent University, Department of Mathematical Physics and
Astronomy, Krijgslaan 281-S9, B-9000 Gent,
Belgium\\E-mail:\email{david.dudal@ugent.be,
henri.verschelde@ugent.be}}

\abstract{We provide a short discussion of the dimension two
condensate $\left\langle A^2\right\rangle$ and its influence on
the infrared behaviour of the gluon propagator in the Landau
gauge. Simultaneously, we pay attention to the issue of Gribov
copies in the Landau gauge. We also briefly discuss a local, gauge
invariant non-Abelian action with mass parameter, constructed from
the dimension 2 operator $F_{\mu\nu} (D^2)^{-1} F_{\mu\nu}$. }

\FullConference{Fifth International Conference on Mathematical
Methods in Physics\\24 -- 28, April 2006\\Rio de Janeiro, Brazil}

\begin{document}
\section{Introduction}
In recent years we have witnessed an intensive research activity
aiming at improving our understanding of the behaviour of Yang-Mills
theories in the low energy (infrared) regime. This is a rather
complicated issue, closely related to the confinement of quarks and
gluons. We consider pure Euclidean $SU(N)$ Yang-Mills theories with
action
\begin{equation}\label{1}
{ S_{YM}}=\frac{1}{4}\int d^4 x\;F_{\mu \nu }^{a}F^{a}_{\mu \nu }\;,
\end{equation}
where $A_{\mu }^{a}$, $a=1,...,N^{2}-1$ is the gauge boson field,
with associated field strength
\begin{equation}\label{3}
F^{a}_{\mu \nu }=\partial _{\mu }A_{\nu }^{a}-\partial _{\nu }A_{\mu
}^{a}+gf^{abc}A_{\mu }^{b}A_{\nu }^{c}\;.
\end{equation}
The theory (\ref{1}) is invariant with respect to the local gauge
transformations
\begin{equation}\label{4}
    \delta A_\mu^a = D_{\mu}^{ab}\omega^b\;,
\end{equation}
with
\begin{equation}\label{5}
    D_{\mu}^{ab}=\partial_\mu\delta^{ab}-gf^{abc}A_\mu^c\;,
\end{equation}
denoting the adjoint covariant derivative. \\\\Thanks to the
asymptotic freedom, the coupling constant of the theory turns out
to be small at very high energies, where perturbation theory then
becomes reliable \cite{Gross:1973id, Politzer:1973fx}. However, in
the infrared region, the coupling constant grows and
nonperturbative effects have to be taken into account. \\\\The
introduction of condensates, i.e. the (integrated) vacuum
expectation value of certain operators, allows one to parametrize
certain nonperturbative effects arising from the infrared sector
of e.g. the theory described by (\ref{1}). Via the Operator
Product Expansion (OPE) (viz. short distance expansion), which is
applicable to local operators, one can relate these condensates to
power corrections which give nonperturbative information in
addition to the perturbatively calculable results. \\\\Possible
causes of nonperturbative effects are given by Gribov ambiguities
which affect the Faddeev-Popov quantization procedure and hence
the propagators \cite{Gribov:1977wm}, the existence of
(topologically) nontrivial field configurations like instantons
\cite{Schafer:1996wv}, etc. These effects are not necessarily
unrelated, as e.g. instantons can contribute to the condensate
$\left\langle\alpha_s F_{\mu\nu}^2\right\rangle$
\cite{Schafer:1996wv}, the Faddeev-Popov operator has zero modes
in an instanton background \cite{Maas:2005qt}, etc. \\\\Although
propagators are not gauge invariant quantities, they are, to some
extent, the simplest Green functions which can be studied from the
analytic point of view. As far as four-dimensional Yang-Mills
theory is considered, this task appears to be very difficult for
more complicated Green functions. \\\\During the last decade,
there has been an intensive activity from the lattice community in
the study of the gluon and ghost propagators. We can thus compare,
at least qualitatively, our theoretical predictions with the
available lattice data. \\\\So far, a certain number of gauges
have been considered extensively from the theoretical as well as
from the lattice point of view. This is the case for the Landau,
Coulomb and maximal Abelian gauges. Throughout this paper, we
shall mainly be interested in the Landau gauge
\begin{equation}\label{landau}
    \p^\mu A_{\mu}^a=0\;.
\end{equation}
In particular, we shall focus on the effect of the dimension two
condensate $\left\langle A^2_{\min}\right\rangle$ on the infrared
behaviour of the gluon propagator, in combination with the effects
arising from a treatment of the Gribov problem.
\section{The dimension two condensate}
The possible existence and relevance of a gauge condensate of
dimension two has been advocated a few years ago by
\cite{Gubarev:2000eu,Gubarev:2000nz}. The idea was put forward that
the gauge invariant, dimension two operator $A^2_{\min}$, obtained
by minimizing $A^2$ along its gauge orbit,
\begin{eqnarray}\label{intro1}
    A^2_{\min}&\equiv&\min_{U\in SU(N)}\frac{1}{VT}\int
    d^4x \left(A_\mu^U\right)^2\;,\nonumber\\
    A_\mu^U&=&UA_\mu U^\dagger+U\p_\mu U^\dagger\;,
\end{eqnarray}
might condense, i.e.
\begin{equation}\label{intro2}
    \left\langle A^2_{\min}\right\rangle\neq0
\end{equation}
This nonvanishing condensate might be important for several reasons:
\begin{itemize}
\item $\left\langle A^2_{\min}\right\rangle$ could
serve as an order parameter for the condensation of monopoles,
relevant for the dual Meissner effect in the dual superconductivity
picture of color confinement \cite{Gubarev:2000eu,Gubarev:2000nz}
\item it could account for certain power corrections in
$\frac{1}{Q^2}$ which have been reported in the study of two and
three point correlation functions, see e.g.
\cite{Boucaud:2001st,RuizArriola:2004en}.
\item it could give rise to a dynamical gluon mass $m_g$ via the OPE
\cite{Kondo:2001nq}, which could be relevant for the dual Meissner
effect \cite{Suzuki:2004dw} as well as for obtaining analytic
estimates of glueball spectra \cite{Cornwall:1981zr}.
\end{itemize}
\section{A closer look at $A^2_{\min}$}
The relevance of the operator $A^2_{\min}$ for Yang-Mills gauge
theories is known since many years. It plays a key role in the study
of the Gribov copies and of the geometrical and topological
properties of the space of gauge orbits
\cite{Dell'Antonio:1989jn,Dell'Antonio:1991xt,Semenov}. It can be
expressed as an infinite series of nonlocal terms, see
\cite{Lavelle:1995ty}, namely
\begin{eqnarray}
A_{\min }^{2} &=&\int d^{4}x\left[ A_{\mu }^{a}\left( \delta _{\mu \nu }-\frac{%
\partial _{\mu }\partial _{\nu }}{\partial ^{2}}\right) A_{\nu
}^{a}-gf^{abc}\left( \frac{\partial _{\nu }}{\partial ^{2}}\partial
A^{a}\right) \left( \frac{1}{\partial ^{2}}\partial {A}^{b}\right)
A_{\nu }^{c}\right] \;+O(A^{4})\;.  \label{nonloc}
\end{eqnarray}
From this expression, one can also check explicitly order by order
in the coupling constant that $A^2_{\min}$ is gauge invariant.
However, it is immediately clear that $A^2_{\min}$ is a highly
nonlocal expression, a fact making it very difficult to handle. In
general, it also falls beyond the applicability of the OPE, which
refers to local operators. \\\\So far, the only possibility to
study the operator $A^2_{\min}$ by analytical tools relied on
choosing the Landau gauge (\ref{landau}). Due to the
transversality condition, all nonlocal terms drop out, so that
$A^2_{\min}$ reduces to the local operator $A^2$,
\begin{equation}\label{drop}
    A^2_{\min}\equiv A^2\mbox{ in the Landau gauge}
\end{equation}
\subsection{Effective potential for the local composite operator }
The operator $A^2$ is an example of a local composite operator
(LCO). In \cite{Verschelde:1995jj}, a method was developed to
construct a sensible effective potential for such operators. This
so-called LCO method was later applied to the case of $A^2$ in the
Landau gauge \cite{Verschelde:2001ia}. We couple the operator $A^2$
to the Yang-Mills action by means of a source $J$,
\begin{eqnarray}\label{m4}
 S_{J}&=&S_{YM}+\int d^4x\left(\frac{1}{2}JA_\mu^aA_\mu^a-\frac{1}{2}\zeta
 J^2\right)\,.
\end{eqnarray}
The last term, quadratic in the source $J$, is necessary to kill the
divergences in vacuum correlators like $\left\langle
A^2(x)A^2(y)\right\rangle$ for $x\to y$, or equivalently in the
generating functional $W(J)$, defined as
\begin{equation}\label{m5}
    e^{-W(J)}=\int[\mbox{fields}] e^{-\int d^4x S_J}\,.
\end{equation}
The presence of the LCO parameter $\zeta$ ensures a homogenous
renormalization group equation for $W(J)$. Its arbitrariness can
be overcome by making it a function $\zeta(g^2)$ of the strong
coupling constant $g^2$, allowing one to fix $\zeta(g^2)$ order by
order in perturbation theory in accordance with the
renormalization group equation
\cite{Verschelde:1995jj,Verschelde:2001ia}. \\\\In order to
recover an energy interpretation, the term $\propto J^2$ can be
removed by employing a Hubbard-Stratonovich transformation
\begin{equation}\label{m6}
    1=\int\sigma
    e^{-\frac{1}{2\zeta}\left(\frac{\sigma}{g}+\frac{1}{2}A^2-\zeta J\right)^2}\,,
\end{equation}
leading to the action
\begin{eqnarray}\label{m7}
S&=&S_{YM}+S_\sigma\,,\nonumber\\
S_\sigma&=&\int
d^4x\left(\frac{\sigma^2}{2g^2\zeta}+\frac{1}{2g^2\zeta}g\sigma
A^2+\frac{1}{8\zeta}(A^2)^2\right)\,.
\end{eqnarray} A key ingredient
for the LCO method is the renormalizability of the operator $A^2$.
It was proven in \cite{Dudal:2002pq} that $A^2$ is renormalizable to
all orders of perturbation theory, making use of the Ward identities
in the presence of the operator $A^2$. In addition, an interesting
identity was proven concerning the anomalous dimension
$\gamma_{A^2}$ of the operator $A^2$, first noticed in
\cite{Gracey:2002yt}. It can be shown that $\gamma_{A^2}$ can be
expressed as a linear combination of the gauge beta function $\beta$
and of the anomalous dimension $\gamma_A$ of the gauge field $A$,
according to the relationship \cite{Dudal:2002pq}
\begin{equation}\label{rel}
    \gamma_{A^2}(a)=-\left(\frac{\beta(a)}{a}+\gamma_A(a)\right)\,,\qquad
    a\equiv\frac{g^2}{16\pi^2}\,.
\end{equation}
Starting from (\ref{m7}) it is possible to calculate the effective
potential $V(\sigma)$. The correspondence $\left\langle
\sigma\right\rangle=-g\left\langle A^2\right\rangle$ consequently
provides evidence for a nonvanishing dimension two gluon condensate
using an effective potential approach, if $\left\langle
\sigma\right\rangle\neq0$. It is clear from (\ref{m7}) that
$\left\langle \sigma\right\rangle\neq0$ also induces an effective
gluon mass. $V(\sigma)$ was calculated to two loop order in
\cite{Verschelde:2001ia,Browne:2003uv}, and a nonvanishing
condensate is favoured as it lowers the vacuum energy. The ensuing
effective gluon mass was found to be a few hundred MeV.

\subsection{Restriction to the Gribov horizon}
In the Landau gauge, it is necessary to restrict the domain of
integration in the Feynman path integral at least to the so-called
Gribov region $\Omega$, whose boundary $\p\Omega$ is the first
Gribov horizon, where the first vanishing eigenvalue of the
Faddeev-Popov operator,
\begin{equation}
\mathcal{M}^{ab}=-\partial _{\mu }\left( \partial _{\mu }\delta
^{ab}+gf^{acb}A_{\mu }^{c}\right) \;,  \label{mm1}
\end{equation}
appears. This restriction is necessary due to the existence of the
Gribov copies, which implies that the Landau condition
(\ref{landau}) does not uniquely fix the gauge
\cite{Gribov:1977wm}. \\\\It has been discussed in
\cite{Zwanziger:1989mf,Zwanziger:1992qr} how this restriction can
be accomplished at the Lagrangian level. More precisely, the
starting Yang-Mills measure in the Landau gauge is given by
\begin{equation}
 d\mu _{\gamma }=DA\delta (\partial_\mu A_\mu^a)\det
(\mathcal{M})e^{-\left( S_{\mathrm{YM}}+\gamma ^{4}H\right) }\;,
\label{m1}
\end{equation}
where
\begin{equation}
H=\int d^{4}xh(x)=g^{2}\int d^{4}xf^{abc}A_{\mu }^{b}\left( \mathcal{M}%
^{-1}\right) ^{ad}f^{dec}A_{\mu }^{e}\;,  \label{m3}
\end{equation}
is the so-called horizon function, which implements the restriction
to the Gribov region $\Omega$. Notice that $H$ is nonlocal. The
parameter $\gamma$, known as the Gribov parameter, has the dimension
of a mass and is not free, being determined by the horizon condition
\begin{equation}
\left\langle h(x)\right\rangle =4\left( N^{2}-1\right) \;,
\label{m4}
\end{equation}
where the expectation value $\left\langle h(x)\right\rangle $ has to
be evaluated with the measure $d\mu _{\gamma }$. To the first order,
the horizon condition (\ref{m4}) reads, in $d$ dimensions,
\begin{equation}
1=\frac{N\left( d-1\right) }{4}g^{2}\int \frac{d^{d}k}{\left( 2\pi
\right) ^{d}}\frac{1}{k^{4}+2Ng^{2}\gamma ^{4}}\;.  \label{mm5}
\end{equation}
This equation coincides with the original gap equation derived by
Gribov for the parameter $\gamma$ \cite{Gribov:1977wm}.
\\\\ Albeit
nonlocal, the horizon function $H$ can be localized through the
introduction of a suitable set of additional fields. The final
action reads
\begin{eqnarray}
S&=&S_{0}-\gamma ^{2}g\int d^{d}x\left( f^{abc}A_{\mu }^{a}\varphi
_{\mu }^{bc}+f^{abc}A_{\mu }^{a}\overline{\varphi }_{\mu
}^{bc}-d(N^2-1)\gamma^4\right)\;,  \label{ll1} \nonumber\\
S_{0} &=&S_{\mathrm{YM}}+\int d^{d}x\;\left( b^{a}\partial_\mu A_\mu^{a}+\overline{c}%
^{a}\partial _{\mu }\left( D_{\mu }c\right) ^{a}\right) \;  \nonumber \\
&+&\int d^{d}x\left( \overline{\varphi }_{\mu }^{ac}\partial _{\nu
}\left(
\partial _{\nu }\varphi _{\mu }^{ac}+gf^{abm}A_{\nu }^{b}\varphi _{\mu
}^{mc}\right) -\overline{\omega }_{\mu }^{ac}\partial _{\nu }\left(
\partial _{\nu }\omega _{\mu }^{ac}+gf^{abm}A_{\nu }^{b}\omega _{\mu
}^{mc}\right)
\right.  \nonumber \\
 &\;&\;\;\;\;\;\;\;\;\;\;\;-g\left.\left( \partial _{\nu }\overline{\omega }_{\mu
}^{ac}\right) f^{abm}\left( D_{\nu }c\right) ^{b}\varphi _{\mu
}^{mc}\right) \;.  \label{l2}
\end{eqnarray}
The fields $\left( \overline{\varphi }_{\mu }^{ac},\varphi _{\mu
}^{ac}\right) $ are a pair of complex conjugate bosonic fields.
Similarly, the fields $\left( \overline{\omega }_{\mu }^{ac},\omega
_{\mu }^{ac}\right) $ are anticommuting. The horizon condition is
equivalent with the demand that the quantum effective action
$\Gamma$ obeys
\begin{equation}\label{hor}
    \frac{\p \Gamma}{\p\gamma^2}=0
\end{equation}
As shown in
\cite{Zwanziger:1989mf,Zwanziger:1992qr,Maggiore:1993wq}, the
resulting local action turns out to be renormalizable to all
orders of perturbation theory. Remarkably, we have been able to
prove that this feature is preserved when the local operator
$A_\mu^2$ is coupled to the Zwanziger action \cite{Dudal:2005na}.
This allows for a simultaneous study of the effects of the Gribov
parameter and condensate $\left\langle A^2\right\rangle$. \\\\A
main consequence of the restriction of the domain of integration
to the Gribov region is the fact that the ghost propagator gets
enhanced in the infrared region. Using the gap equation arising
from the horizon condition, one finds that
\cite{Gribov:1977wm,Zwanziger:1989mf,Zwanziger:1992qr,Sobreiro:2004us,Dudal:2005na,Gracey:2005cx}
\begin{equation}\label{enh}
    \left\langle
    c^a\overline{c}^b\right\rangle_p\sim\frac{\delta^{ab}}{p^4}\mbox{ for }
    p^2\to 0
\end{equation}
This enhancement remains valid in the presence of $\left\langle
A^2\right\rangle$ \cite{Sobreiro:2004us,Dudal:2005na}. The
infrared enhancement of the ghost propagator in the Landau gauge
has also been observed from lattice simulations
\cite{Maas:2006qw,Furui:2006rx} or solutions of the
Schwinger-Dyson equations
\cite{Watson:2001yv,Lerche:2002ep,Zwanziger:2001kw}. \\\\The
Gribov restriction and $\left\langle A^2\right\rangle$ also affect
the gluon propagator in a nontrivial fashion, more precisely one
finds \cite{Sobreiro:2004us,Dudal:2005na}
\begin{equation}\label{glp}
\left\langle
    A_\mu^a A_\nu^b \right\rangle_p=\delta^{ab}\left(\delta_{\mu\nu}-\frac{p_\mu p_\nu}{p^2}\right)\frac{p^2}{p^4+m^2p^2+2g^2N\gamma^4}
\end{equation}
It is worth noticing that a propagator like (\ref{glp}) is not
new, as it has been considered already some time ago by Stingl in
order to solve the Schwinger-Dyson equations \cite{Stingl:1985hx}.
It induces an infrared suppressed gluon propagator, a fact in
qualitative agreement with lattice \cite{Maas:2006qw} and
Schwinger-Dyson results
\cite{Watson:2001yv,Lerche:2002ep,Zwanziger:2001kw}. Let us also
mention that (\ref{glp}) violates spectral positivity, giving an
indication that the gauge bosons are unphysical particles
\cite{Dudal:2005na}

\section{Beyond the Landau gauge}
It is unclear what the role of $A^2_{\min}$ might be in other
gauges. This is a very difficult question, without any answer at
the moment. Due to the severe nonlocality in the expression
(\ref{nonloc}), it seems that explicit calculations outside the
Landau gauge are almost prohibitive, as a localization procedure
looks quite hopeless. \\\\Nevertheless, in several other gauges,
we have shown that other dimension two, renormalizable, local
operators exist. We generalized the LCO method and showed that
these operators condense and give rise to a dynamical gluon mass,
see Table 1 and \cite{Dudal:2004rx,Dudal:2003by,Dudal:2003gu}.
\begin{table}
\begin{center}
\begin{tabular}{|c|c|}
  \hline
  Gauge&LOperator\\
  \hline\hline
  linear covariant &   $\frac{1}{2}A_\mu^a A_\mu^a$\\
  Curci-Ferrari & $\frac{1}{2}A_\mu^a
  A_\mu^a+\alpha\overline{c}^ac^a$\\
  maximal Abelian & $\frac{1}{2}A_\mu^{\beta} A_\mu^{\beta}+\alpha\overline{c}^{\beta}c^{\beta}$ \\
  \hline
\end{tabular}
\caption{Gauges and their renormalizable dimension two operator}
\end{center}
\end{table}
In the maximal Abelian gauge, it was found that only the
off-diagonal gluons $A_\mu^\beta$ acquire a dynamical mass, a fact
qualitatively consistent with the lattice results from
\cite{Amemiya:1998jz,Bornyakov:2003ee}. \\\\We have been able to
make some connection between the various gauges and their
dimension two operators by constructing renormalizable
interpolating gauges and operators
\cite{Dudal:2004rx,Dudal:2005zr}. However, these operators are
explicitly gauge dependent, hence there does not seem to exist a
clear relation with the gauge invariant operator $A^2_{\min}$.
\\\\Recently, the issue of Gribov copies has also been addressed in
the maximal Abelian gauge \cite{Capri:2005tj,Capri:2006cz}.

\section{Another gauge invariant dimension two operator}
Recently, we have considered the fact that, perhaps, another gauge
invariant dimension two operator might be of some relevance. We
took a look at
\begin{equation}\label{op}
\mathcal{O}\equiv\frac{1}{VT}\int d^{4}xF_{\mu \nu }^{a}\left[
\left( D^{2}\right) ^{-1}\right] ^{ab}F_{\mu \nu }^{b}\;.
\end{equation}
This operator was already introduced by Jackiw and Pi during their
analysis of a dynamical mass generation in 3-dimensional gauge
theories \cite{Jackiw:1995nf}. \\\\We can add the operator
(\ref{op}) to the Yang-Mills action as a mass term via
\begin{equation}
S_{YM}+S_{\mathcal{O}}\;,  \label{ymop}
\end{equation}
with
\begin{equation}
S_{\mathcal{O}}=-\frac{m^{2}}{4}\int d^{4}xF_{\mu \nu }^{a}\left[
\left( D^{2}\right) ^{-1}\right] ^{ab}F_{\mu \nu }^{b}\;.
\label{massop}
\end{equation}
As we have discussed in \cite{Capri:2005dy}, the action (\ref{ymop})
can be localized by introducing a pair of complex bosonic
antisymmetric tensor fields, $\left( B_{\mu \nu
}^{a},\overline{B}_{\mu \nu }^{a}\right) $, and a pair of complex
anticommuting antisymmetric tensor fields, $\left( \overline{G}_{\mu
\nu }^{a},G_{\mu \nu }^{a}\right) $, belonging to the adjoint
representation, according to which
\begin{eqnarray}
e^{-S_{\mathcal{O}}}&=&\int D\overline{B}DBD\oG DG\exp \left[
-\left( \frac{1}{4}\int d^{4}x\overline{B}_{\mu \nu }^{a}D_{\sigma
}^{ab}D_{\sigma }^{bc}B_{\mu \nu }^{c}\right.\right.\nonumber\\&-&\left.\left.\frac{1}{4}\int {%
d^{4}x}\overline{G}_{\mu \nu }^{a}D_{\sigma }^{ab}D_{\sigma
}^{bc}G_{\mu
\nu }^{c}+\frac{im}{4}\int d^{4}x\left( B-\overline{B%
}\right) _{\mu \nu }^{a}F_{\mu \nu }^{a}\right)
\right]\;.\label{loc2}
\end{eqnarray}
Doing so, we obtain a classical local action which reads
\begin{equation}
S_{YM}+S_{BG}+S_{m}\;,  \label{action1}
\end{equation}
where
\begin{eqnarray}
S_{BG} &=&\frac{1}{4}\int d^{4}x\left( \overline{B}_{\mu \nu
}^{a}D_{\sigma }^{ab}D_{\sigma }^{bc}B_{\mu \nu
}^{c}-\overline{G}_{\mu \nu }^{a}D_{\sigma
}^{ab}D_{\sigma }^{bc}G_{\mu \nu }^{c}\right) \;,  \nonumber \\
S_{m} &=&\frac{im}{4}\int d^{4}x\left( B-\overline{B}\right) _{\mu
\nu }^{a}F_{\mu \nu }^{a}\;,  \label{actions2}
\end{eqnarray}
which is left invariant by the gauge transformations
\begin{eqnarray}
\delta A_{\mu }^{a} &=&-D_{\mu }^{ab}\omega ^{b}\;,  \nonumber \\
\delta B_{\mu \nu }^{a} &=&gf^{abc}\omega ^{b}B_{\mu \nu
}^{c}\;,\;\; \delta \overline{B}_{\mu \nu }^{a} =gf^{abc}\omega
^{b}\overline{B}_{\mu \nu }^{c}\;,
\nonumber \\
\delta G_{\mu \nu }^{a} &=&gf^{abc}\omega ^{b}G_{\mu \nu
}^{c}\;,\;\; \delta \overline{G}_{\mu \nu }^{a} =gf^{abc}\omega
^{b}\overline{G}_{\mu \nu }^{c}\;. \label{gtm}
\end{eqnarray}
In order to discuss the renormalizability of (\ref{action1}), we
relied on the method introduced by Zwanziger in
\cite{Zwanziger:1989mf,Zwanziger:1992qr} to discuss the
renormalizability of the nonlocal horizon function (\ref{m3}).
Instead of using (\ref{action1}) with $m$ coupled to the composite
operators $B_{\mu\nu}^aF_{\mu\nu}^a$ and $B_{\mu\nu}^aF_{\mu\nu}^a$,
we introduce 2 suitable external sources $V_{\rho\sigma\mu\nu}$ and
$\overline{V}_{\rho\sigma\mu\nu}$ and replace $S_m$ by
\begin{equation}
\frac{1}{4}\int d^{4}x\left( V_{\sigma \rho \mu \nu
}\overline{B}_{\sigma \rho }^{a}F_{\mu \nu
}^{a}-\overline{V}_{\sigma \rho \mu \nu }B_{\sigma \rho }^{a}F_{\mu
\nu }^{a}\right) \;. \label{rs}
\end{equation}
At the end, the sources $V_{\sigma \rho \mu \nu }(x)$,
$\overline{V}_{\sigma \rho \mu \nu }(x)$ are required to attain
their physical value, namely
\begin{equation}
\overline{V}_{\sigma \rho \mu \nu }\Big|_{\mathrm{phys}}=V_{\sigma \rho \mu \nu }%
\Big|_{\mathrm{phys}}=-\frac{im}{2}\left( \delta _{\sigma \mu
}\delta _{\rho \nu }-\delta _{\sigma \nu }\delta _{\rho \mu }\right)
\;, \label{ps}
\end{equation}
so that (\ref{rs}) reduces to $S_m$ in the physical limit. \\\\We
assume the linear covariant gauge fixing, implemented through
\begin{equation}
S_{gf}=\int d^{4}x\left( \frac{\alpha }{2}b^{a}b^{a}+b^{a}\partial
_{\mu }A_{\mu }^{a}+\overline{c}^{a}\partial _{\mu }D_{\mu
}^{ab}c^{b}\right) \;, \label{lg}
\end{equation}
In \cite{Capri:2005dy}, we wrote down a list of symmetries enjoyed
by the action
\begin{equation}
S_{YM}+S_{BG}+S_{gf}\;,
\end{equation}
i.e. in absence of the sources. Let us only mention here the BRST
symmetry, generated by the nilpotent transformation $s$ given by
\begin{eqnarray}
sA_{\mu }^{a} &=&-D_{\mu }^{ab}c^{b}\;,\;\;  sc^{a} =\frac{g}{2}f^{abc}c^{a}c^{b}\;,  \nonumber \\
sB_{\mu \nu }^{a} &=&gf^{abc}c^{b}B_{\mu \nu }^{c}+G_{\mu \nu
}^{a}\;,\;\; s\overline{B}_{\mu \nu }^{a}
=gf^{abc}c^{b}\overline{B}_{\mu \nu }^{c}\;,
\nonumber\\
sG_{\mu \nu }^{a} &=&gf^{abc}c^{b}G_{\mu \nu }^{c}\;,\;\;
s\overline{G}_{\mu \nu }^{a} =gf^{abc}c^{b}\overline{G}_{\mu \nu
}^{c}+\overline{B}_{\mu
\nu }^{a}\;,  \nonumber \\
s\overline{c}^{a} &=&b^{a} \;,\;\; sb^{a} =0\;,\;\;s^{2} =0\;.
\label{bi}
\end{eqnarray}
It turns out that one can introduce all the necessary external
sources in a way consistent with the starting symmetries. This
allows to write down several Ward identities by which the most
general counterterm is restricted using the algebraic
renormalization formalism \cite{Capri:2005dy}. After a very
cumbersome analysis, it turns out that the action (\ref{action1})
must be modified to
\begin{eqnarray}
  S_{phys} &=& S_{cl} +S_{gf}\;,\label{completeaction}
  \end{eqnarray}
  with
  \begin{eqnarray}
  S_{cl}&=&\int d^4x\left[\frac{1}{4}F_{\mu \nu }^{a}F_{\mu \nu }^{a}+\frac{im}{4}(B-\overline{B})_{\mu\nu}^aF_{\mu\nu}^a
  +\frac{1}{4}\left( \overline{B}_{\mu \nu
}^{a}D_{\sigma }^{ab}D_{\sigma }^{bc}B_{\mu \nu
}^{c}-\overline{G}_{\mu \nu }^{a}D_{\sigma }^{ab}D_{\sigma
}^{bc}G_{\mu \nu
}^{c}\right)\right.\nonumber\\
&-&\left.\frac{3}{8}%
m^{2}\lambda _{1}\left( \overline{B}_{\mu \nu }^{a}B_{\mu \nu
}^{a}-\overline{G}_{\mu \nu }^{a}G_{\mu \nu }^{a}\right)
+m^{2}\frac{\lambda _{3}}{32}\left( \overline{B}_{\mu \nu
}^{a}-B_{\mu \nu }^{a}\right) ^{2}\right.\nonumber\\&+&\left.
\frac{\lambda^{abcd}}{16}\left( \overline{B}_{\mu\nu}^{a}B_{\mu\nu}^{b}-\overline{G}_{\mu\nu}^{a}G_{\mu\nu}^{b}%
\right)\left( \overline{B}_{\rho\sigma}^{c}B_{\rho\sigma}^{d}-\overline{G}_{\rho\sigma}^{c}G_{\rho\sigma}^{d}%
\right) \right]\;, \label{completeactionb}\label{lcg}
\end{eqnarray}
in order to have renormalizability to all orders of perturbation
theory. We notice that we had to introduce a new invariant quartic
tensor coupling $\lambda^{abcd}$, subject to the generalized Jacobi
identity
\begin{equation}\label{jacobigen}
    f^{man}\lambda^{mbcd}+f^{mbn}\lambda^{amcd}+f^{mcn}\lambda^{abmd}+f^{mdn}\lambda^{abcm}=0\,,
\end{equation}
and the symmetry constraints
\begin{eqnarray}
\lambda^{abcd}=\lambda^{cdab} \;, \nonumber \\
\lambda^{abcd}=\lambda^{bacd} \;, \label{abcd}
\end{eqnarray}
as well as two new mass couplings $\lambda_1$ and $\lambda_3$.
Without the new couplings, i.e. when $\lambda_1\equiv0$,
$\lambda_3\equiv0$, $\lambda^{abcd}\equiv0$, the previous action
would not be renormalizable. We refer to
\cite{Capri:2005dy,Capri:2006ne} for all the details. We also
notice that the novel fields $B_{\mu\nu}^a$,
$\overline{B}_{\mu\nu}^a$, $G_{\mu\nu}^a$ and
$\overline{G}_{\mu\nu}^a$  are no longer appearing at most
quadratically. As it should be expected, the classical action
$S_{cl}$ is still gauge invariant w.r.t. (\ref{gtm}). \\\\The BRST
transformation (\ref{bi}) no longer generates a symmetry of the
gauge fixed action $S_{phys}$. However, we are able to define a
natural generalization of the usual BRST symmetry that does
constitute an invariance of the gauge fixed action
(\ref{completeaction}). Indeed, after inspection, one shall find
that
\begin{eqnarray}
  \widetilde{s}S_{phys} &=& 0\;, \nonumber\\
  \widetilde{s}^2 &=& 0\;,
\end{eqnarray}
with
\begin{eqnarray}
\widetilde{s} A_{\mu }^{a} &=&-D_{\mu }^{ab}c ^{b}\;,\;\;  \widetilde{s} c^{a} =\frac{g}{2}f^{abc}c^ac ^{b}\;,  \nonumber \\
 \widetilde{s} B_{\mu \nu }^{a} &=&gf^{abc}c ^{b}B_{\mu \nu }^{c}\;,\;\;\widetilde{s} \overline{B}_{\mu \nu }^{a} =gf^{abc}c
^{b}\overline{B}_{\mu \nu }^{c}\;,
\nonumber \\
\widetilde{s} G_{\mu \nu }^{a} &=&gf^{abc}c ^{b}G_{\mu \nu
}^{c}\;,\;\;\widetilde{s} \overline{G}_{\mu \nu }^{a} =gf^{abc}c
^{b}\overline{G}_{\mu
\nu }^{c}\;,\nonumber\\
\widetilde{s}\overline{c}^{a} &=&b^a\;,\;\; \widetilde{s} b^{a}
=0\;.
 \label{brst3}
\end{eqnarray}
In \cite{Capri:2005dy,Capri:2006ne}, we also calculated explicitly
various renormalization group equations to two loop order,
confirming the renormalizability at the practical level. Various
consistency checks are at our disposal in order to establish the
reliability of these results, e.g. the gauge parameter independence
of the anomalous dimension of gauge invariant quantities or the
equality of others, in accordance with the output of the Ward
identities in \cite{Capri:2005dy}. Furthermore, we proved in
\cite{Capri:2006ne} the equivalence of the model
(\ref{completeaction}) with the ordinary Yang-Mills theory in the
case that $m\equiv0$, making use of the nilpotent transformation
\begin{eqnarray}\label{ss}
\delta_s B_{\mu\nu}^a &=& G_{\mu\nu}^a \;,\qquad \delta_s G_{\mu\nu}^a =0 \;,\nonumber\\
\delta_s \overline{G}_{\mu\nu}^a &=& \overline{B}_{\mu\nu}^a \;,
\qquad \delta_s \overline{B}_{\mu\nu}^a = 0 \;,\nonumber\\ \delta_s
(\textrm{rest})&=&0\;,
\end{eqnarray}
which generates a ``supersymmetry'' of the action
$S_{phys}^{m\equiv0}$.

\section*{Acknowledgments}
The Conselho Nacional de Desenvolvimento Cient\'{i}fico e
Tecnol\'{o}gico (CNPq-Brazil), the Faperj, Funda{\c{c}}{\~{a}}o de
Amparo {\`{a}} Pesquisa do Estado do Rio de Janeiro, the SR2-UERJ
and the Coordena{\c{c}}{\~{a}}o de Aperfei{\c{c}}oamento de Pessoal
de N{\'\i}vel Superior (CAPES) are gratefully acknowledged for
financial support.

\end{document}